 \documentstyle[aps]{revtex}
 
\begin{document}
\title{Nonlinear Transport in a Quantum Point Contact due to Soft Disorder
Induced Coherent Mode Mixing }
\draft
\author{A.M. Zagoskin and R.I. Shekhter}
\address{Department of Applied Physics, Chalmers University of Technology and
University of G\"{o}teborg, S-412 96, G\"{o}teborg, Sweden}
\date{\today}
\maketitle
\begin{abstract}
We show that the coherent   mixing of different transverse modes, due to
forward scattering of carriers by soft impurity- or boundary potentials
leads to a nonlinear, asymmetric current response of quantum point contacts
(QPC).
The oscillating contribution to the current is sensitive both to  driving
voltage  and to gate voltage  in direct analogy to the electrostatic
Aharonov-Bohm effect.

  Our calculations are in a good agreement with  recent  experimental data
showing  small-scale conductivity nonlinearities  and asymmetry    in QPC.
\end{abstract}


It is well known that the quantization of the electron's transverse momentum in
quantum point contacts (QPC) - structures, defined in the 2D electron gas of
GaAs-AlGaAs interfaces by the  electrostatic potential of the gate electrodes -
is responsible for their nonlinear behaviour both in regard to the gate
voltage, $V_{g}$, and the driving voltage, $V_{sd}$. The nonlinearities are due
to effective turning on-off the conducting modes inside the contact when
$eV_{g}$\cite{vanWees,Beenakker,GLKhSh1988}, or
$eV_{sd}$\cite{Glazman,First,Alex,Pepper} exceeds the interlevel spacing of
transverse electronic modes in the contact region,
 $\Delta E_{\perp} \simeq E_{F}/N \simeq 1 {\rm meV}$
(N-number of open modes). They were observed experimentally, as differential
conductance quantized steps vs. $V_{g}$ \cite{vanWees,Beenakker} or as extrema
of the conductance vs. $V_{sd}$ derivative \cite{Pepper}.

The quantization effects in QPC proved to be stable in regard to the potential
variations in the system (e.g., due to charged impurities). The reason is that
backscattering is needed in order to violate the quantization, while
scattering on soft potentials present in the system cannot produce a sufficient
momentum change \cite{GLKhSh1988,Glazman/Jonson}. This explains the success of
an adiabatic model of QPC \cite{GLKhSh1988} in spite of significant mode
mixture due to forward scattering inside the system \cite{Chao}.

 Nevertheless, these results don't  exhaust all the  possibilities,  and in
recent experiments \cite{Danmark} a more complicated picture was observed,
which cannot be explained in the framework of the existing approach.

The nonlinearity and  asymmetry ( sensitivity to the current direction ) of QPC
response was observed at driving voltages as small as 0.01 mV, both directly in
measurements of the differential resistance vs. current  and via ac current
rectification.  The effect was clearly seen at 0.3 K and on increasing the
temperature till 4.2 K it was smeared very similarly to the smearing of the
transconductance, $dG/dV_{g}$. Resistance variation was of order 20 $\Omega$,
on the background of QPC resistance of several k$\Omega$ (QPC conductance
ranged from 1 to 10 $(2e^{2}/h)$-quanta).

The purpose of the present paper is to demonstrate that the scattering on soft
potentials can play a major role in the ballistic transport  through
microconstrictions due to coherent mixing of the transverse modes, their phases
being defined by $V_{sd}$ and $V_{g}$, in a direct analogy to the electrostatic
Aharonov-Bohm effect.

The effect exists in a QPC with at least two scatterers (impurities, defects,
nonadiabaticities of the contact's shape etc). In a system with only soft
scatterers  a crucial role in its development is played by the mechanism of
indirect  backscattering  suggested by Laughton et al. \cite{Davies} plays .
(It is related to the possibility for a carrier to be switched from the
transport mode to non-propagating one even  by the forward (small-angle)
scattering.)

We will show, that   a qualitative understanding of the measured oscillations
and asymmetry of the contact's conductance is achieved. The characteristic
scales of the oscillations vs. $V_{sd}, V_{g}$ are in good accordance with the
experimental data as well.

The model we consider is  an adiabatically smooth 2D channel connecting two
equilibrium reservoirs, with two scatterers inside.
The wave function of an electron with energy E in some point of our system can
be expanded over  WKB eigenfunctions for transporting modes \cite{GLKhSh1988}

\begin{eqnarray}
\psi^{\alpha}(x,y;E) = \sum_{m,\alpha}a_{m}^{\alpha}\chi_{m}^{\alpha}(x,y;E);
\label{exp}\\
\chi_{m}^{\alpha}(x,y;E) = \sqrt{p_{\infty}(E)/p_{m\parallel}(E,x)} \exp\left(
i/\hbar \int_{(-1)^{\alpha}\cdot\infty}^{x} dx' p_{m\parallel}(x';E) \right)
\phi_{m}(y; x).
\end{eqnarray}

In this expression $\phi_{m}(y; x)$ is a transverse eigenfunction with
eigenvalue  $E_{m\perp}(x)=\left(p_{m\perp}(x;E)\right)^{2}/2m^{*}$; index
$\alpha = 0,1$ numbers the waves propagating from the right ($\alpha=0$) or
left ($\alpha=1$) reservoir.

The phase gained by a  partial wave between the scatterers A,B is

\begin{equation}
\sigma_{n}(E) = 1/\hbar\int_{x_{B}}^{x_{A}} dx' p_{n\parallel}(x'; E) =
1/\hbar\int_{x_{B}}^{x_{A}} dx'\sqrt{2m\left(E-E_{n\perp}(x)-e\Phi(x)\right)},
\end{equation}
and is evidently dependent on the mode number, electron energy and absolute
value of the external electrical potential in the interscatterer region,
$\Phi(x)$. Phase differences between  different partial waves are thus
sensitive to $\Phi(x)$, in a  direct analogy to the electrostatic Aharonov-Bohm
effect, and so is an interference contribution to the current.

The current through the system is written in a standard way \cite{Imry}:

\begin{eqnarray}
I(V_{sd})
= e/(\pi\hbar)\int dE \left[ n_{F}(E-\mu_{R})-n_{F}(E-\mu_{L})\right] {\rm
Tr}\left(\hat{T}^{0}(E)^{\dag}\hat{T}^{0}(E)\right). \label{tok}
\end{eqnarray}
Here $\hat{T}^{0}(E)$ is the transfer-matrix for the particle with energy $E$
incident  from the right reservoir:
$
\Psi^{0}_{out}(E) = \hat{T}^{0}(E)\Psi^{0}_{in}(E).
$
N-dimensional vector columns $\Psi^{\alpha}_{in,out}$   contain the expansion
coefficients $a_{m; in,out}^{\alpha}$ Eq.(~\ref{exp}) (amplitudes of transverse
modes in in- and outcoming  waves  incident from the right or left reservoir
respectively). {\em N} is a number of conducting modes in the channel.

The scatterers are described   by unitary (2N$\times$2N) S-matrices
\cite{Matrix}. If we introduce the phase gain matrix,
$
\hat{U}_{ij} =  \delta_{ij}\exp(i\sigma_{j}(E)),
$
then the transfer-matrix is expressed through $\hat{U}$ and elements of
S-matrices of scatterers as follows:
\begin{eqnarray}
\hat{T}^{0}(E) = \hat{t}^{0}_{B}\hat{U}\hat{t}^{0}_{A} +
\hat{t}^{0}_{B}\hat{U}\hat{r}^{0}_{A}\hat{U}
\hat{r}^{1}_{B}\hat{U}\hat{t}^{0}_{A} + \dots  = \hat{t}^{0}_{B}\left(\hat{1}
-\hat{U}\hat{r}^{0}_{A} \hat{U}\hat{r}^{1}_{B}\right)^{-1}
\hat{U}\hat{t}^{0}_{A}.
\end{eqnarray}
Here $\hat{t}^{\alpha}_{A,B}, \hat{r}^{\alpha}_{A,B}$ are (N$\times$N)-matrices
 which describe  the transmission (reflection) to the left ($\alpha=0$), or to
the right ($\alpha=1$) (see e.g. \cite{Matrix}) . Summation of higher-order
terms, corresponding to    multiple passages of the wave  through the system,
allows us to obtain  conductance resonances like the ones predicted for QPCs
with sharp edges \cite{SzaferStone,Kirczenow,ZagKulik}, which could contribute
to small-scale nonlinearities in the system.   This contribution is  negigible
in our case because of two reasons. First,  these resonances are due to direct
backscattering into the same mode and were never obtained in more realistic
models with contacts having a smooth  shape. Second,  the period of the
corresponding nonlinearities is much smaller than that of  oscillations due to
intermode mixing, and they  surely would have vanished at the temperature of
the experiment.

Keeping thus only  the first term of the series, which contains effects of
intermode mixing, we find that the current  is given by  (~\ref{tok}) with
\begin{eqnarray}
 {\rm Tr} \left(\hat{T}^{0}(E)^{\dag}\hat{T}^{0}(E)\right) =
\sum_{j}\sum_{k} \left(\hat{t}_{A}^{0}\hat{t}_{A}^{0\dag}\right)_{jk}
\left(\hat{t}_{B}^{0\dag}\hat{t}_{B}^{0}\right)_{kj}
\exp\left(i(\sigma_{j}(E)-\sigma_{k}(E))\right). \label{current}
\end{eqnarray}

  The effect we are interested in is contained in phase-sensitive off-diagonal
terms of the double sum.

The exact form of the matrices $\hat{t}$  cannot be determined  without  using
some specific model of a scatterer, be it impurity,   contact-bank interface or
something else. We are not here concerned with these  poorly controllable
details, and concentrate on the nonlinear  conductivity of QPC as a function of
$V_{sd}$ and $V_{g}^{eff}$. Nevertheless some general remarks are to be made.

Unitarity condition imposes constraints upon the transition and reflection
submatrices. In the general case \cite{Matrix}
\begin{eqnarray}
\left(\hat{t}_{A}^{0}\hat{t}_{A}^{0\dag}\right) =
\hat{u}_{1}(\hat{1}+\hat{\lambda}_{A})^{-1}\hat{u}_{1}^{\dag};
\left(\hat{t}_{B}^{0\dag}\hat{t}_{B}^{0}\right) =
\hat{u}_{2}^{\dag}(\hat{1}+\hat{\lambda}_{B})^{-1}\hat{u}_{2},
\end{eqnarray}
where $\hat{u}_{1,2}$ are arbitrary unitary (N$\times$N)-matrices, and
$\hat{\lambda}$ is a diagonal (N$\times$N)-matrix with nonnegative elements.
$\hat{\lambda}$ equals to zero if and only if the reflection submatrix   is
zero, i.e., if the direct  backscattering is absent.

In the latter case both $\left(\hat{t}_{L}^{0\dag}\hat{t}_{L}^{0}\right) =
\left(\hat{t}_{R}^{0}\hat{t}_{R}^{0\dag}\right) = \hat{1}$, and interference
terms are exactly zero, {\em no matter how strong the mode mixing is}. Thus, as
long as number of open modes is not changed ($V_{sd}<\Delta V_{\perp}$),
conductance quantization is not affected by finite driving voltage at all.
This is consistent with  numerical calculations by Brataas and Chao
\cite{Chao}, where mode mixing was shown to be unimportant for conductance
quantization in the linear response regime. We see that this is a consequence
of unitarity.

Direct backscattering on soft potentials,  present in QPC, is usually very
small \cite{Glazman/Jonson}. Nevertheless  Laughton et al \cite{Davies}
suggested a more effective backscattering mechanism: {\em indirect}
backscattering  through localized modes, i.e. standing waves trapped in bulges
in the channel. They showed numerically that such modes really appear in
realistic QPC due to overlap of long-range impurity potentials (therefore this
process can not be described within  Born approximation), and that in their
presence backscattering is much more intensive, and exact conductance
quantization is violated.

In order to demonstrate this mechanism in our matrix model, regard an extremal
case: two {\em nonreflecting} scatterers (with $\hat{\lambda}=0$) inside the
bulge (see Fig.\ref{fig1}). Bottlenecks are modelled by two projection
operators, $\hat{P},$ allowing  to pass only $N_{P}<N$  transverse modes.  We
come to expression (\ref{tok}) with

 \begin{eqnarray}
{\rm Tr}\left(\hat{T}^{0}(E)^{\dag}\hat{T}^{0}(E)\right) =
{\rm Tr} \left\{\left(\hat{1}-\hat{M}\right)^{-1\dag}
\left(\hat{t}^{0\dag}_{B}\hat{P}\hat{t}^{0}_{B}\right)
\left(\hat{1}-\hat{M}\right)^{-1} \hat{U} \left(\hat{t}^{0}_{A}
\hat{P}\hat{t}^{0\dag}_{A}\right) \hat{U}^{\dag} \right\},
\end{eqnarray}
where $\hat{M}=\hat{U}\hat{t}^{0}_{A}(\hat{1}-\hat{P})\hat{t}^{1}_{A}\hat{U}
\hat{t}^{1}_{B}(\hat{1}-\hat{P})\hat{t}^{0}_{B}.$

If $\hat{P} \neq \hat{1}$  interference terms are present, since generally
neither $\hat{t}^{0}_{R}\hat{P}\hat{t}^{0\dag}_{R}$, nor
$\hat{t}^{0\dag}_{L}\hat{P}\hat{t}^{0}_{L}$    is proportional to the unit
matrix.

  From now then we are concerned only with nonlinearities in $I(V_{sd},V_{g})$
dependence. $N$ stands  here  for a number of  propagating modes. Suppose  that
the chemical potential of the left reservoir is constant, $\mu_{L}=E_{F}$,
while $\mu_{R}=E_{F}+eV_{sd}$\cite{footnote}.

Neglecting an energy dependence of scattering matrix at Fermi level, we can
then write the dimensionless nonlinear contribution  to current, $i_{osc}\equiv
I_{osc}\cdot(\pi\hbar/e)/(AE_{F})$ as

\begin{eqnarray}
i_{osc}(v,u) = \int d\epsilon
\left[n_{F}(\epsilon-1-v)-n_{F}(\epsilon-1)\right]
\cos\left(\sigma_{1}(\epsilon;v,u)-\sigma_{2}(\epsilon;v,u)\right).
\label{Dimless}
\end{eqnarray}
Here  $A \leq 1/2$ is a numerical constant dependent on details   of the system
configuration, and the  dimensionless variables $\epsilon=E/E_{F},
\epsilon_{j\perp}(x)=E_{j\perp}(x)/E_{F}, v=eV_{sd}/E_{F},
u=eV_{g}^{eff}/E_{F}, n=W/(\lambda_{F}/2),$ and $l=L/\lambda_{F}$ are
introduced \cite{footnote2}.

It follows from (\ref{Dimless}), that the effect can be observed at
temperatures of order or less than its characteristic energy scale (being an
order of magnitude less than the intermode separation, $\Delta E_{\perp}$),
that is, a few Kelvin.

In the experiment \cite{Danmark} the rectified voltage was measured, as a
response to ac current (Fig.\ref{fig2}(a)),
\begin{equation}
V_{rect}(I_{0},V_{g}) = 1/2\pi \int_{0}^{2\pi} d\phi I_{0}\sin\phi \:\:
R(I_{0}\sin\phi, V_{g}).
\end{equation}
This quantity includes effects of both nonlinearity and asymmetry of the QPC
response.

In our approach, though, it is easier to calculate the rectified  current:
\begin{equation}
i_{rect}(v_{0},u) = 1/2\pi \int_{0}^{2\pi} d\phi v_{0}\sin\phi \:\:
G(v_{0}\sin\phi ,u),
\label{irec}
\end{equation}
and then define an effective rectified voltage by
\begin{equation}
v_{rect}^{eff}(i_{0},u)/(i_{0}R_{N}(u)) = i_{rect}(v_{0},u)/(v_{0}G_{N}(u)),
\label{effvg}
\end{equation}
where $G_{N}(u) = R_{N}^{-1}(u) = N(u)e^{2}/\pi\hbar$ is the electrical
conductance of an ideal QPC with $N$ open modes.

The results of numerical  calculations of $v_{rect}^{eff}(u)$ are given in
Fig\ref{fig2}. In order to simplify calculations, the contact was shaped as a
straight channel with impenetrable walls, of (dimensionless) width $n$. The
influence of gate voltage was reduced to coordinate-independent lift of the
potential  instead of changing the channel's width; this agrees with
experimental situation \cite{Pepper}.  All the driving voltage   was supposed
to drop on the right end of the channel. In order to  evaluate an  effect of
multiple mode mixing, in the following calculations we have summed up all
possible interference terms with Gaussian weightes. For example, $\delta G =
\sum_{i=1}^{N}\sum_{j=1}^{i-1} \delta G_{ij} \exp(-(i-j)^{2}/2s).$ Weight
factor are introduced to reduce contribution of less probable processes with
scattering to larger angles ($s$ being a mixing parameter).

 A qualitative likeness between the experimental and theoretical curves is
evident. The period of $v_{rect}^{eff}(u)$ as function of the effective gate
voltage  in our calculations corresponds to ($\simeq 0.1$ mV), and  is thus of
the same order as the characteristic potential shift in the contact's centre
determined in \cite{Danmark} ( $\simeq 100 \mu$V).  Thus the model gives  a
proper order-of-magnitude description  of the effect. The magnitude of observed
oscillations (less than $10^{-4}$ of total conductance) can be provided, e.g.,
by mode mixing on the contact-bank interfaces \cite{Peter} or impurity
potentials \cite{Davies}.

 In conclusion, we have shown that coherent mode  mixing in quantum point
contacts leads to a new type of nonlinearities in this system, which differs
from   nonlinearity due to conductance quantization    and have  an energetic
scale $\simeq 0.1$ mV.  These oscillations develop on expence of conductance
quantization, and  may occur even in systems with soft potentials. They are
sensitive to the current direction, thus providing an explanation for  the
observed asymmetry of the current-voltage curve in QPC. The  magnitude of the
oscillations is defined by the backscattering rate in QPC governed in the  case
of soft scattering potentials by indirect backscattering through the localized
modes.  The effect can be observed at temperatures    of order $\Delta V/k_{B}
\simeq 1$K.

The effect is a probable cause of recently observed small-scale nonlinearities
and asymmetry of I-V curve in QPC.

The sensitivity of the effect to  weak potential changes in the vicinity of the
contact suggests that the slow changes in time of the impurity configuration
near QPC would reveal as a low frequency conductance noise. This may provide an
explanation for the random telegraph noise observed   in the system as well.

\acknowledgements

We are grateful to P.E.Lindelof and M.Persson for aquanting us
with their recent results, and to them and  I.E.Aronov, Yu.M.Galperin,
M.Jonson, I.V.Krive,  and S.N.Rashkeev for many fruitful discussions.

\begin{figure}
\caption{  Interference effects in a QPC with two nonreflecting scatterers
inside a bulge. Localized modes in the bulge appear because only $N_{P}$   of
$N$ availiable  transverse modes can pass the bottlenecks.
{\em Inset:} Indirect backscattering  by a soft potential via the localized
mode. 1)An electron is scattered from the propagating mode {\em a} to the
localized mode {\em b} (standing wave) with a small momentum change. 2)It is
scattered from the localized mode to the mode {\em c} propagating in opposite
direction,  also with a small momentum change. }
\label{fig1}
\end{figure}

\begin{figure}
\caption{Rectified voltage vs. gate voltage curves.
(a) Experimental curve from Taboryski et al. [10].  The solid and broken lines
correspond to different values of ac current. Inset shows the corresponding
conductance vs. gate voltage  characteristic. The step width is of order 0.1 V,
the maximal QPC conductance correspomds to 10 propagating modes.
(b) Effective rectified voltage vs. effective gate voltage, calculated
according to (13),(14). The contact length $l=100/2\pi$, width $n=10.1$,
$v_{0}=0.03$,  Gaussian mode mixing is supposed with $s=1$. The  idealized
conductance $G_{N}(u) = (2e^{2}/h)N(u)$  is shown as well. The number of open
modes, $N(u)$,   corresponds to the experimental situation, figure (a).  }
\label{fig2}
\end{figure}


\begin{references}
\bibitem{footnote} As we have pointed earlier, the phases of partial waves feel
the absolute value of the electrical potential between the scatterers.
Therefore the oscillations exist independently on the details of the voltage
drop (of course, their shape and (a)symmetry depend on it). Since the positions
of the relevant scatterers vs. the potential profile in given QPC are random,
the asymmetric case is more probable. We discuss here the limiting case; its
generalization to arbitrary degree of asymmetry is trivial.
\bibitem{footnote2}$V_{g}^{eff}$ is the effective gate voltage, i.e., the
electic potential change induced in the centre of the contact  by the gate
potential. It is usually one-two orders of magnitude less than actual voltage
on the gate electrode.
\end{references}
\end{document}